\documentclass[11pt,a4paper]{article}
\usepackage{amsmath,amssymb}
\usepackage{epsfig,graphicx}
\usepackage{tabularx} 
\usepackage{subfigure}
\usepackage{enumurate}
\graphicspath{{//Users/mkh/Documents/PhD_Thesis/UoMThesis/figures/Sky/} {//Users/mkh/Documents/Sky_paper/Fit/}{//Users/mkh/Documents/Sky_prod/}{//Users/mkh/Documents/Sky_paper/Min/}{//Users/mkh/Documents/Sky_paper/Spin/}}

\newcommand{\no}{\nonumber}

\begin{document}

\title{\bf Isospinning Skyrmions}
\author{R.~A.~Battye$^a$\footnote{{\bf e-mail}: richard.battye@manchester.ac.uk}, M.~Haberichter$^b$\footnote{{\bf e-mail}: m.haberichter@kent.ac.uk},
S.~Krusch$^{b}$\footnote{{\bf e-mail}: s.krusch@kent.ac.uk}\\
$^a$ \small{\em Jodrell Bank Centre for Astrophysics} \\
\small{\em University of Manchester, Manchester M13 9PL, U.K.}\\
$^b$ \small{\em School of Mathematics, Statistics and Actuarial Science} \\
\small{\em University of Kent, Canterbury CT2 7NF, U.K.}\\
}

\author{M.~Haberichter$^{a,b}$\footnote{{\bf e-mail}: m.haberichter@kent.ac.uk}\\
$^a$ \small{\em School of Mathematics, Statistics and Actuarial Science} \\
\small{\em University of Kent, Canterbury CT2 7NF, U.K.}\\
$^b$ \small{\em Department of Applied Mathematics and Theoretical Physics} \\
\small{\em University of Cambridge, Wilberforce Road, Cambridge CB3 0WA, U.K.}\\
}

\date{}
\maketitle

\begin{abstract}
In the Skyrme model atomic nuclei are modelled as quantized soliton solutions in a nonlinear field theory of pions. The mass number is given by the conserved topological charge $B$ of the solitons. Conventionally, Skyrmions are semiclassically quantized within the rigid body approach. In this approach Skyrmions are effectively treated as rigid rotors in space and isospace that is it is assumed that Skyrmions do not deform at all when they spin and isospin. This approximation resulted in qualitative and encouraging quantitative agreement with experimental nuclear physics data. In this talk, we point out that the theoretical agreement could be further improved by allowing classical Skyrmion solutions to deform as they spin and isospin. As a first step towards a better understanding of how nuclei can be approximated by classically spinning and isospinning soliton solutions, we study how classical Skyrmion solutions of topological charges $B=1-4,8$ deform when classical isospin is added. 
\end{abstract}

\section{Introduction}

Skyrmions are topological soliton solutions of the Skyrme model \cite{Skyrme:1961vq,Skyrme:1962vh} -- a nonlinear effective field theory of pions. They can be used to describe nucleons and nuclei \cite{Adkins:1983ya,Adkins:1983hy}, with an identification between soliton and baryon numbers.  
The allowed quantum states of light atomic nuclei of mass numbers 1 to 8, 10, 12 have been determined  \cite{Manko:2007pr,Battye:2009ad} and in most cases correct spin, parity and isospin quantum numbers for the ground states and various excited states have been obtained. Recently,  Skyrmion solutions of baryon number 12 have been used to reproduce the rotational excitations of Carbon-12 \cite{Lau:2014baa}. In particular, the recently experimentally observed rotational band of the Hoyle state can be understood quantitatively.   

Nevertheless, the vast majority of the nuclear spectra  calculated from the Skyrme model are based on the rigid body quantization \cite{Adkins:1983ya,Adkins:1983hy}. This is effectively a study of rigidly spinning Skyrmion solutions,  that is  any deformations and symmetry changes that might be due to centrifugal effects are neglected. The allowed spin and isospin quantum numbers are determined from the symmetries of the static classical Skyrme soliton solutions for vanishing spin and isospin, see for example Refs. \cite{Irwin:1998bs,Krusch:2002by,Krusch:2005iq}. In particular, for even mass numbers, the allowed quantum states for each Skyrmion and their excitation energies \cite{Battye:2009ad} are in reasonable agreement with experimental observed nuclei states. However, the Skyrme model's predictions for odd mass numbers \cite{Manko:2007pr} are far less accurate. For example, to obtain the correct physical ground states of Lithium-7 and Beryllium-7 nuclei, the charge-7 Skyrmion has to be deformed significantly \cite{Manko:2006dr}. The icosahedrally symmetric Skyrmion solution which is the lowest energy solution \cite{Battye:2001qn} with baryon number 7 and for vanishing spin and isospin is found to be far too symmetric to model the lowest allowed spin states of the ${}^7\text{Li}/{}^7\text{Be}$ isospin doublet. 

\begin{figure}[!htb]
\centering
\includegraphics[totalheight=6.0cm]{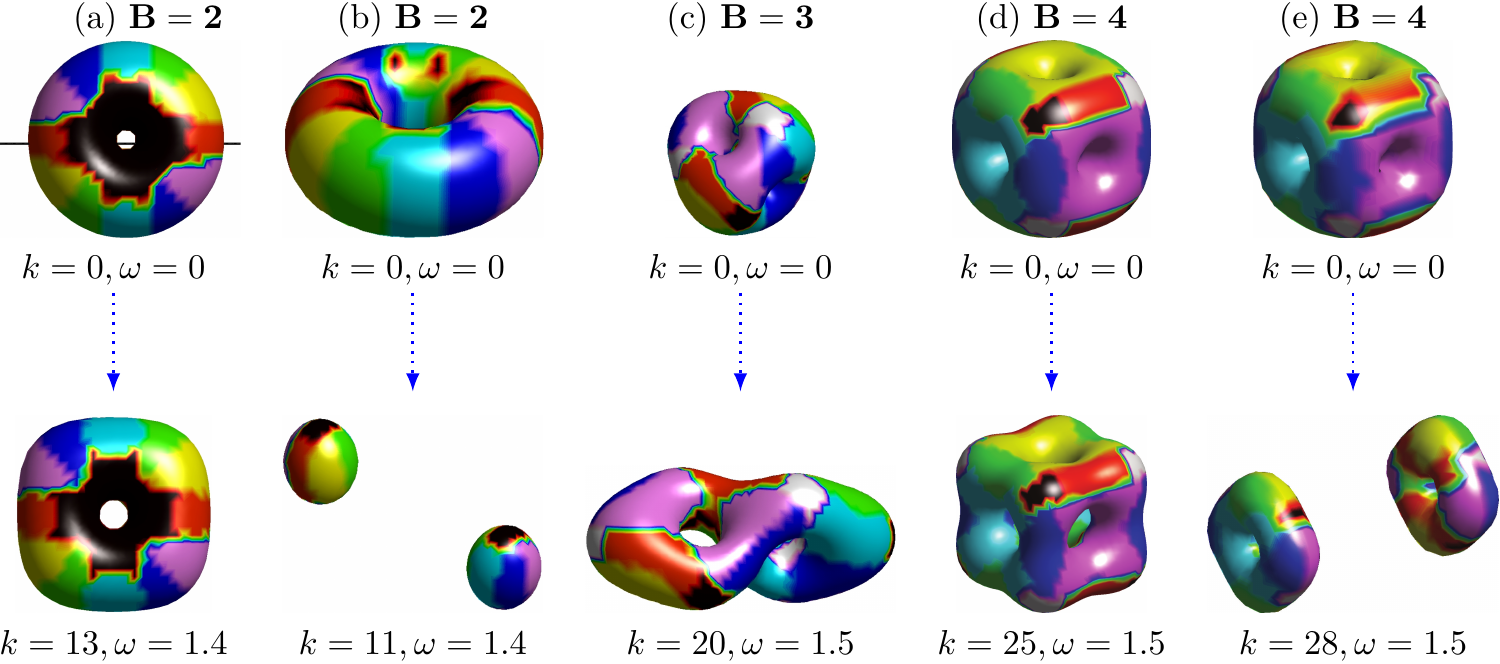}
\caption{Baryon density isosurfaces of $B=2-4$ Skyrmion solutions with pion mass $\mu=1.5$ and as a function of isospin $K$ and angular frequency $\omega$. Note that the isospin $K$ is given in units of $4\pi$, i.e. we define $k = K/4\pi$. The isosrotation axes are chosen to be: (a) $\boldsymbol{\widehat{K}}=(0,1,0)$, (b) $\boldsymbol{\widehat{K}}=(0,0,1)$,  (c)  $\boldsymbol{\widehat{K}}=(0,0,1)$, (d) $\boldsymbol{\widehat{K}}=(0,0,1)$, (e) $\boldsymbol{\widehat{K}}=(0,1,0)$.}
\label{Fig_B1B8_Sky}
\end{figure}

It has been pointed out by several authors \cite{Rajaraman:1985ty,Battye:2005nx,Houghton:2005iu,Fortier:2008yj,Manton:2011mi} that one very promising way to improve the Skyrme model's agreement with experimental nuclear physics data is to allow Skyrmion solutions to deform when they spin and isospin. This can change the symmetries of the solutions and might result in different allowed quantum states \cite{Manko:2007pr}. Indeed, allowing Skyrmions to deform axially symmetrically as they spin resulted for charge-1 and charge-2 Skyrmion solutions in a significant reduction of their rotational energy contribution \cite{Battye:2005nx,Houghton:2005iu,Fortier:2008yj}. Furthermore, it revealed that there does not exist a spinning charge-1 Skyrmion solution that models both the nucleon and the delta resonance if the pion mass parameter $m_\pi$ of the Skyrme model is set to its experimental value. Finally, recent numerical investigations of isospinning soliton solutions in other models of the Skyrme family (e.g. baby Skyrmions \cite{Battye:2013tka,Halavanau:2013vsa}, Hopf solitons \cite{Battye:2013xf,Harland:2013uk}) beyond the rigid body approximation demonstrated that the geometrical shape and energies can be significantly affected when centrifugal effects are taken into account. 

A systematic study of classically spinning and isospinning Skyrmion solutions beyond the rigid body approximation is still lacking. However, in our  recent article \cite{Battye:2014qva} we performed numerical full field simulations of isospinning Skyrmion solutions with baryon numbers $B=1-4, 8$, without imposing any spatial symmetries. Our investigations can be seen as a first step towards a better
understanding of classically spinning Skyrmions and their possible applications in nuclear and particle physics.

The talk is based on our recent publication \cite{Battye:2014qva} and organized as follows. In Section~\ref{Sec_Sky_Iso} we briefly sketch our approach to the minimization problem of constructing classically isospinning soliton solutions in the Skyrme Model. In Section~\ref{Sec_Numerics} we point out our main findings on isospinning Skyrme solitons with topological charges $B=1-4,\,8$. A brief summary and conclusion of our results is given in Section~\ref{Sec_Con}.

\section{Isospinning Skyrmion Solutions}\label{Sec_Sky_Iso}

\begin{figure}[htb]
\centering
\includegraphics[totalheight=6.0cm]{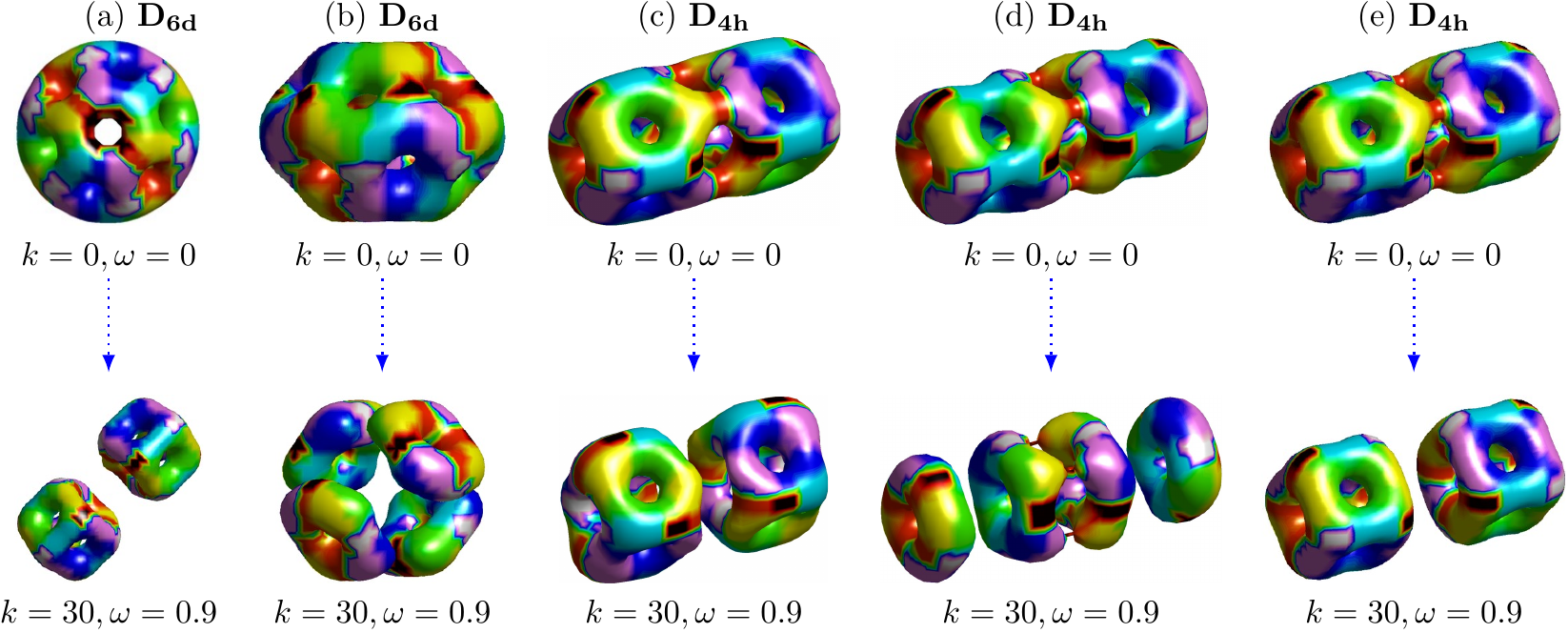}
\caption{Baryon density isosurfaces of $D_{6d}$ and $D_{4h}$ symmetric $B=8$ Skyrmion solutions with pion mass $\mu=1$ and as a function of isospin $K$. The isosrotation axes are chosen to be: (a) $\boldsymbol{\widehat{K}}=(0,0,1)$, (b) $\boldsymbol{\widehat{K}}=(0,1,0)$,  (c)  $\boldsymbol{\widehat{K}}=(1,0,0)$, (d) $\boldsymbol{\widehat{K}}=(0,1,0)$, (e) $\boldsymbol{\widehat{K}}=(0,0,1)$.}
\label{Fig_B8_Sky}
\end{figure}

Restricting to static field configurations, we can define the Skyrme model by its energy functional 
\begin{align}
M_B&=\int\left\{\left(\partial_i\boldsymbol{\phi}\cdot\partial_i\boldsymbol{\phi}\right)+\tfrac{1}{2}\Big[\left(\partial_i\boldsymbol{\phi}\cdot\partial_i\boldsymbol{\phi}\right)^2-\left(\partial_i\boldsymbol{\phi}\cdot\partial_j\boldsymbol{\phi}\right)^2\Big]+2\mu^2\left(1-\sigma\right)\right\}\,\text{d}^3x\,,
\label{Sky_mass}
\end{align}
where $\mu$ is a dimensionless pion mass parameter  and we use the sigma model notation. We collect the scalar meson field $\sigma$ and the pion isotriplet $\boldsymbol{\pi}=(\pi_1,\pi_2,\pi_3)$ together in a four component unit vector $\boldsymbol{\phi}=(\sigma,\boldsymbol{\pi})$. To ensure fields have finite potential energy and a well-defined integer degree $B$ the Skyrme field $\boldsymbol{\phi}$ has to approach the vacuum configuration at spatial infinity, that is $\boldsymbol{\phi}\rightarrow(1,0,0,0)$ as $|x|\rightarrow \infty$. 
Therefore, the domain can be formally compactified to a 3-sphere $S^3_{\text{space}}$ and the Skyrme field $\boldsymbol{\phi}$ is then given by a mapping $S^3_{\text{space}}\rightarrow  S^3_{\text{iso}}$ labelled by the topological invariant $B=\pi_3(S^3)\in\mathbb{Z}$. We display in the first rows of Fig.~\ref{Fig_B1B8_Sky} and \ref{Fig_B8_Sky} baryon density isosurfaces for the Skyrmion solutions of minimal energy for baryon numbers $B=1-4, 8$ and with pion mass value $\mu$. For $B=1$ the soliton solutions have been found to have spherical symmetry, axial symmetry for $B=2$, platonic symmetry for $B=3,4$ and dihedral symmetry for $B=8$. For baryon number $B=8$, there exist two very different Skyrme configurations of comparable energy \cite{Battye:2006na} -- one with $D_{4h}$ and the other with $D_{6d}$ symmetry. 

\begin{table}[htb]
\caption{Skyrmions of baryon number $B=1-4,\,8$. We list the energies $M_B$, the energy per baryon $M_B/B$, the diagonal elements of the inertia tensors $U_{ij},V_{ij},W_{ij}$ and the symmetry group $G$ of the Skyrme solitons. Note that energies $M_B$ are given in units of $12\pi^2$ and that the mass parameter is chosen to be $\mu=1$. The calculated  configurations correspond to global energy minima for given baryon number $B$. For $B=8$ we are unable to decide within the limits of our numerical accuracy which configuration is of lower energy.}
\begin{tabularx}{\textwidth}{cXXXXXXXXXXXc}
\hline\hline
 $B$& $G$  & $M_B$   &$M_B/B$&$U_{11}$ &$U_{22}$ &$U_{33}$&$V_{11}$ &$V_{22}$ &$V_{33}$ &$W_{11}$ &$W_{22}$ &$W_{33}$\\
\hline
1& $O(3)$ &$1.415$  & 1.415& 47.5 &      47.5  &    47.5  &47.5 &47.5 &47.5&47.5 &47.5&47.5\\
2& $D_{\infty h}$  &$2.720$ &1.360  & 97.0 & 97.0&68.9 &153.8 &153.8&275.4 & 0.0 &0.0 & 137.7 \\
3& $T_d$           &$3.969$& 1.323 &124.1 &124.1 &124.1 &402.8 & 402.8 &402.8 &85.2 & 85.2 &85.2\\
4& $O_h$  &$5.177$ &1.294 & 148.2 & 148.2 & 177.4 & 667.6 &667.6 &667.6 & 0.0 & 0.0 & 0.3 \\
8 &$D_{6d}$&$10.235$&1.279& 296.3& 296.3 & 285.2 & 2261.4 &2261.4 &3036.3 &0.1 & 0.1 &137.9\\
&$D_{4h}$& 10.235&  1.279 & 298.4 & 292.1  &  326.9 &4093.9 & 4094.8 & 1381.3 & 0.1 &0.0 &0.1\\\hline\hline
\end{tabularx}
\label{Tab_Stat_sky_low}
\end{table}

Note that the classical soliton mass (\ref{Sky_mass}) is expressed in terms of \emph{Skyrme units}, that is the energy and length units are given by $F_\pi/4e$ and $2/eF_\pi$, respectively. Here $e$ is a dimensionless parameter and $F_\pi$ is the pion decay constant. The dimensionless pion mass $\mu$ is proportional to the tree-level pion mass $m_\pi$, explicitly $\mu=2m_\pi/eF_\pi$. Conventionally, the Skyrme parameters $e$ and $F_\pi$ are determined by fixing the energies of rigidly spinning spin 1/2 and spin 3/2 Skyrmions  to the masses of the nucleon and delta resonance, assuming the experimental value $m_\pi=138\,\text{MeV}$ for the pion mass \cite{Adkins:1983ya,Adkins:1983hy}. This approach results in the standard values $F_\pi=108\,\text{MeV}$, $e=4.84$ and $\mu=0.526$. Throughout this talk we consider pion values $\mu$ between 0.5 and 2. These pion mass values are motivated by the observations in Refs.~\cite{Manko:2007pr,Battye:2009ad,Battye:2005nx,Battye:2006tb} that larger pion mass values (in particular, $\mu>0.526$) yield improved results when applying the Skyrme model to nuclear physics, while there are also studies which lead to lower values of $\mu$ see e.g. Ref.~\cite{Kopeliovich:2004pd}. In fact, the same reference \cite{Kopeliovich:2004pd} also considers the effect of a sixth order Skyrme term, which now has become important in so-called Bogomolny-Prasad-Sommerfield Skyrme models \cite{Adam:2013wya,Adam:2013tda}.

The construction of spinning and isospinning Skyrmion solutions requires the computation of the inertia tensors. The isospin ($U_{ij}$), spin ($V_{ij}$) and mixed ($W_{ij}$) inertia tensors are given explicitly by the integrals \cite{Battye:2014qva,Kopeliovich:2001yg,Lau:2014sva}
\begin{subequations}\label{Sky_Inertia}
\begin{align}
U_{ij}&=2\int\Bigg\{\left(\pi_d\pi^d\delta_{ij}-\pi_i\pi_j\right)\left(1+\partial_k\boldsymbol{\phi}\cdot\partial_k\boldsymbol{\phi}\right)-\epsilon_{ide}\epsilon_{jfg}\left(\pi^d \partial_k\pi^e\right)\left(\
\pi^f\partial_k\pi^g\right)\Bigg\}\,{\text{d}}^3x\,,\label{U_phi_not}\\
V_{ij}&=2\int\epsilon_{ilm}\epsilon_{jnp}x_lx_n\Bigg(\partial_m\boldsymbol{\phi}\cdot\partial_p\boldsymbol{\phi}-\left(\partial_k\boldsymbol{\phi}\cdot\partial_m\boldsymbol{\phi}\right)\left(\partial_k\
\boldsymbol{\phi}\cdot\partial_p\boldsymbol{\phi}\right)\no\\
&\quad\quad\quad\quad\quad\quad\quad\quad+\left(\partial_k\boldsymbol{\phi}\cdot\partial_k\boldsymbol{\phi}\right)\left(\partial_m\boldsymbol{\phi}\cdot\partial_p\boldsymbol{\phi}\right)\
\Bigg)\,{\text{d}}^3x\label{V_phi_not}\,,\\
W_{ij}&=2\int\epsilon_{jlm}x_l\Bigg(\epsilon_{ide}\pi^d\partial_m\pi^e\left(1+\partial_k\boldsymbol{\phi}\cdot\partial_k\boldsymbol{\phi}\right)-\left(\partial_k\boldsymbol{\phi}\cdot\partial_m\boldsymbol{\phi}\right)\left(\epsilon_{ifg}\pi^f\partial_k\pi^g\right)\Bigg)\,{\text{d}}^3x\,.
\end{align}
\end{subequations}
We list in Table~\ref{Tab_Stat_sky_low} the energies and the diagonal elements of the inertia tensors $U_{ij},V_{ij},W_{ij}$ for Skyrmions with baryon numbers $B=1-4, 8 $ of symmetry group $G$. We find that in each case the off-diagonal entries are small and are of the order of $10^{-2}$ times the diagonal entries or less. In the following, all inertia tensor elements will be rounded to one decimal place.

Uniformly isospinning soliton solutions in Skyrme models are obtained by solving one of the following equivalent variational problems \cite{Harland:2013uk} for $\boldsymbol{\phi}$:
\begin{enumerate}
\item[(1)] Extremize the pseudoenergy functional $F_\omega\left(\boldsymbol{\phi}\right)=M_B-\frac{1}{2}\omega_iU_{ij}\omega_j$ for fixed angular frequency $\boldsymbol{\omega}$\,,
\item[(2)] Extremize the Hamiltonian $H=M_B+\frac{1}{2}K_iU_{ij}^{-1}K_j$ for fixed isospin $K_i=U_{ij}\omega_j$.
\end{enumerate}

In this talk, we will use a hybrid of approach (1) and (2). We fix the isospin ${\bf K}$ to be constant. Then we consider the energy
\begin{equation}
\label{EOmega0}
E = M_B + \frac{1}{2} \omega_i U_{ij} \omega_j\,,
\end{equation}
which implies
\begin{equation}
\label{omegaK}
K_i = U_{ij} \omega_j\,.
\end{equation}
The corresponding body-fixed spin angular momenta are given by
\begin{equation}
\label{omegaL}
L_i = -W_{ij}^T \omega_j = -W_{ij} U_{jk}^{-1} K_k\,.
\end{equation}
Hence, in our approach, if $W_{ij}$ is nonzero, then the configuration will obtain \emph{classical} spin. We will discuss this further in section~\ref{Sec_Numerics}. We could now express the energy (\ref{EOmega0}) as a function of ${\bf K}$ and then minimize the energy $E$. However, it is more convenient to calculate $\omega$ using (\ref{omegaK}) and then minimize the pseudoenergy
\begin{equation}
\label{F}
F_\omega = M_B - \frac{1}{2} \omega_i U_{ij} \omega_j\,.
\end{equation}
Since we only fix ${\bf K}$ but not ${\bf L},$ the value of $\omega$ is not conserved during the minimization. Hence for each step, we recalculate $\omega$ using (\ref{omegaK}).

\section{Numerical Results}\label{Sec_Numerics}

We construct stationary isospinning soliton solutions by numerically solving the energy minimization problem formulated in Section~\ref{Sec_Sky_Iso}. Note that spinning Skyrmions with zero pion mass radiate away their energy, see Ref.~\cite{Schroers:1993yk} for a detailed discussion. For pion mass $\mu>0$ stationary solutions exist up to an angular frequency $\omega_{{\rm crit}} = \mu$. At $\omega_{\text{crit}}$ the values of the energy and angular momentum are finite, and therefore, the corresponding angular momenta $K_{\text{crit}}$ (and $L_{\text{crit}}$) is also finite, see Ref.~\cite{Battye:2005nx}. The situation is different for baby Skyrmions where energy and moment of inertia diverge at $\omega_{{\rm crit}}$ \cite{Battye:2013tka,Halavanau:2013vsa,Piette:1994mh} for $\mu<1$. From the point of view of numerics, this behavior is challenging. For $\omega < \omega_{\text{crit}}$ the problem is well-posed, whereas for $\omega>\omega_{\text{crit}}$ the solutions become oscillatory which is difficult to detect in a finite box. Physically, this corresponds to pion radiation, and the fact that stationary solutions do not exist. Numerically, we can find energy minimizers for $\omega > \omega_{\text{crit}}$ but this is an artefact of the finite box approximation. By convention, throughout this talk, when displaying inertia tensors and energies as a function of isospin, we will cut our graphs at the critical isospin value $K_{\text{crit}}$.  

The numerical energy minimization calculations are performed with a second order gradient flow algorithm with a friction term included. During the energy relaxation kinetic energy is removed periodically by setting $\dot{\boldsymbol{\phi}}=0$ at all grid points every 50th timestep. We use the rational map ansatz \cite{Houghton:1997kg} to generate initial Skyrme field configuration of given topological charge $B$ and symmetry group $G$. For each isospin value $\boldsymbol{|K|}$ the damped field evolution algorithm converges to a minimum of the pseudoenergy functional (\ref{F}). Minimal energy solutions at lower $|\boldsymbol{K}|$ are used as initial conditions for higher $|\boldsymbol{K}|$. 

\begin{figure}[!htb]
\centering
\includegraphics[totalheight=3.0cm]{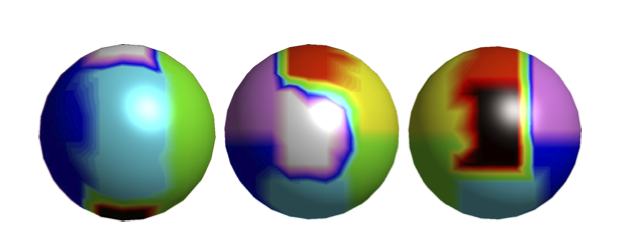}
\caption{Three different views of the baryon isodensity of a $B=1$ Skyrmion. The orientation in isospace is visualized using the field colouring scheme given in Refs.~\cite{Manton:2011mi,Battye:2006na,Feist:2012ps}.}
\label{Colourscheme_Manton}
\end{figure}

Note that we orientate the Skyrmion solutions in isospace such that their principal axes are aligned with the chosen isorotation axes. We visualize the different orientations of Skyrme solitons using Manton's and Sutcliffe's field colouring scheme \cite{Manton:2011mi}. We illustrate the colouring for a $B=1$ Skyrmion solution in Fig.~\ref{Colourscheme_Manton}: The points where the normalised pion isotriplet $\widehat{\boldsymbol{\pi}}$ takes the values $\widehat{\pi}_1=\widehat{\pi}_2=0$ and $\widehat{\pi}_3=+1$ are shown in white and those where $\widehat{\pi}_1=\widehat{\pi}_2=0$ and $\widehat{\pi}_3=-1$ are coloured black. The red, blue and green regions indicate where $\widehat{\pi}_1+i\widehat{\pi}_2$ takes the values $1$, $e^{2\pi i/3}$, $e^{4\pi i/3}$, respectively and the associated complementary colors in the RGB colour scheme (cyan, yellow and magenta) show the segments where $\widehat{\pi}_1+i\widehat{\pi}_2=-1, e^{5\pi i/3}, e^{\pi i/3}$. 

For topological charges $B=1-4$, our numerical simulations are carried out on regular, cubic grids of $(200)^3$ grid points with a lattice spacing $\Delta x=0.1$ and time step size $\Delta t=0.01$. For charge-8 Skyrmion solutions, we use cubic grids containing $(201)^3$ lattice points and with the same lattice spacing  $\Delta x=0.1$. The finite difference scheme used is fourth order accurate in the spatial derivatives. More details on the numerical implementation can be found in Refs.~\cite{Battye:2001qn,Battye:2014qva}.

\subsection{Lower Charge Skyrmions ($1\le B\le 4$)}

In this section, we summarize our findings \cite{Battye:2014qva} on isospinning Skyrme solitons of topological charges $B=1-4$.  We investigate how classical isospin affects the geometrical shape and energy of Skyrmions. For each Skyrme configuration we perform numerical simulations of isospinning Skyrme solitons for all possible choices of  the isospin axis. Numerical results are shown for mass values $\mu=0.5,1,1.5$ and $2$. 

\subsubsection{$B=1$}

\begin{figure}[!htb]
\centering
\subfigure[\,Mass-Spin relationship ]{\includegraphics[totalheight=5.5cm]{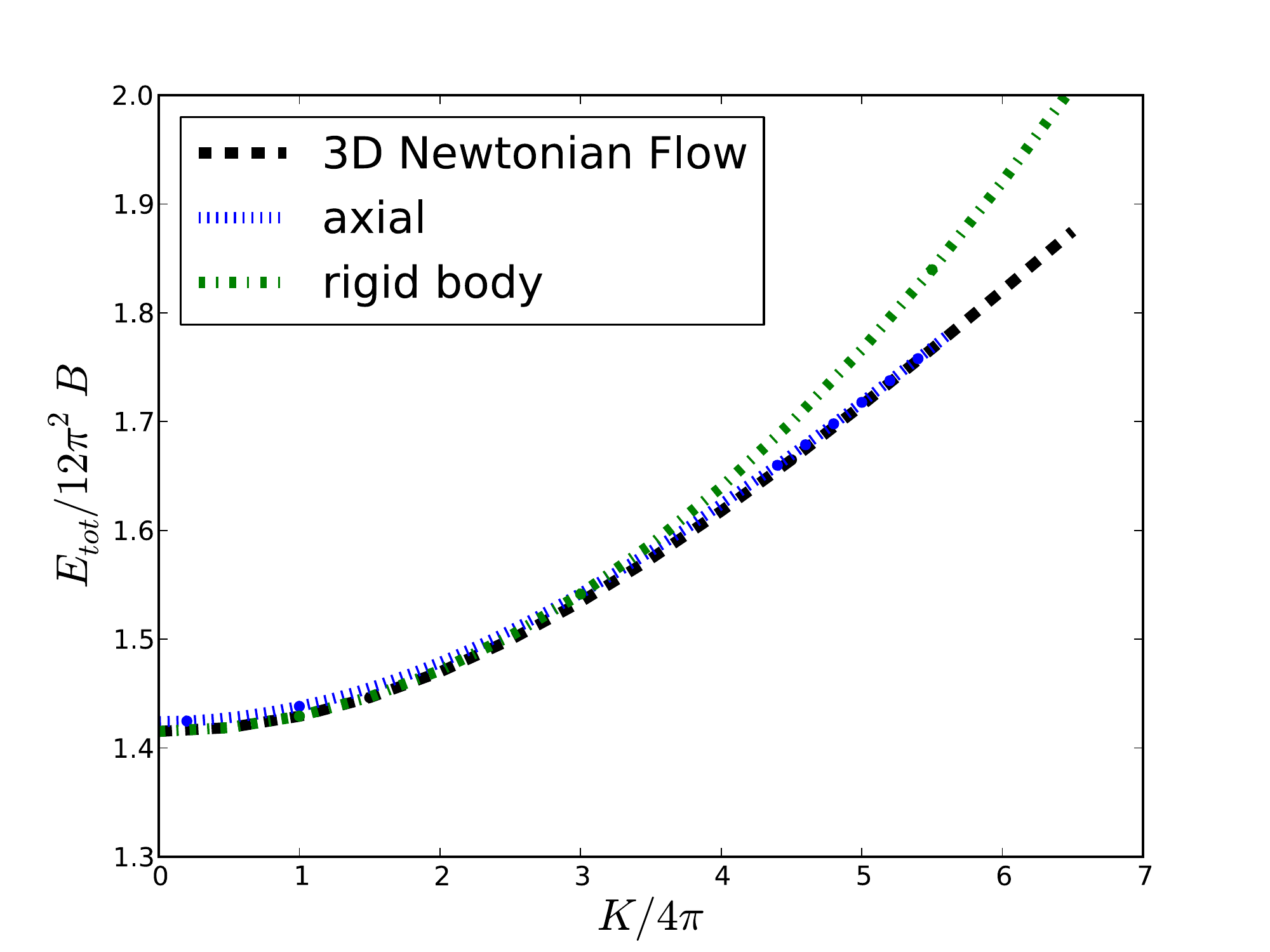}}
\subfigure[\,Inertia-Spin relationship]{\includegraphics[totalheight=5.5cm]{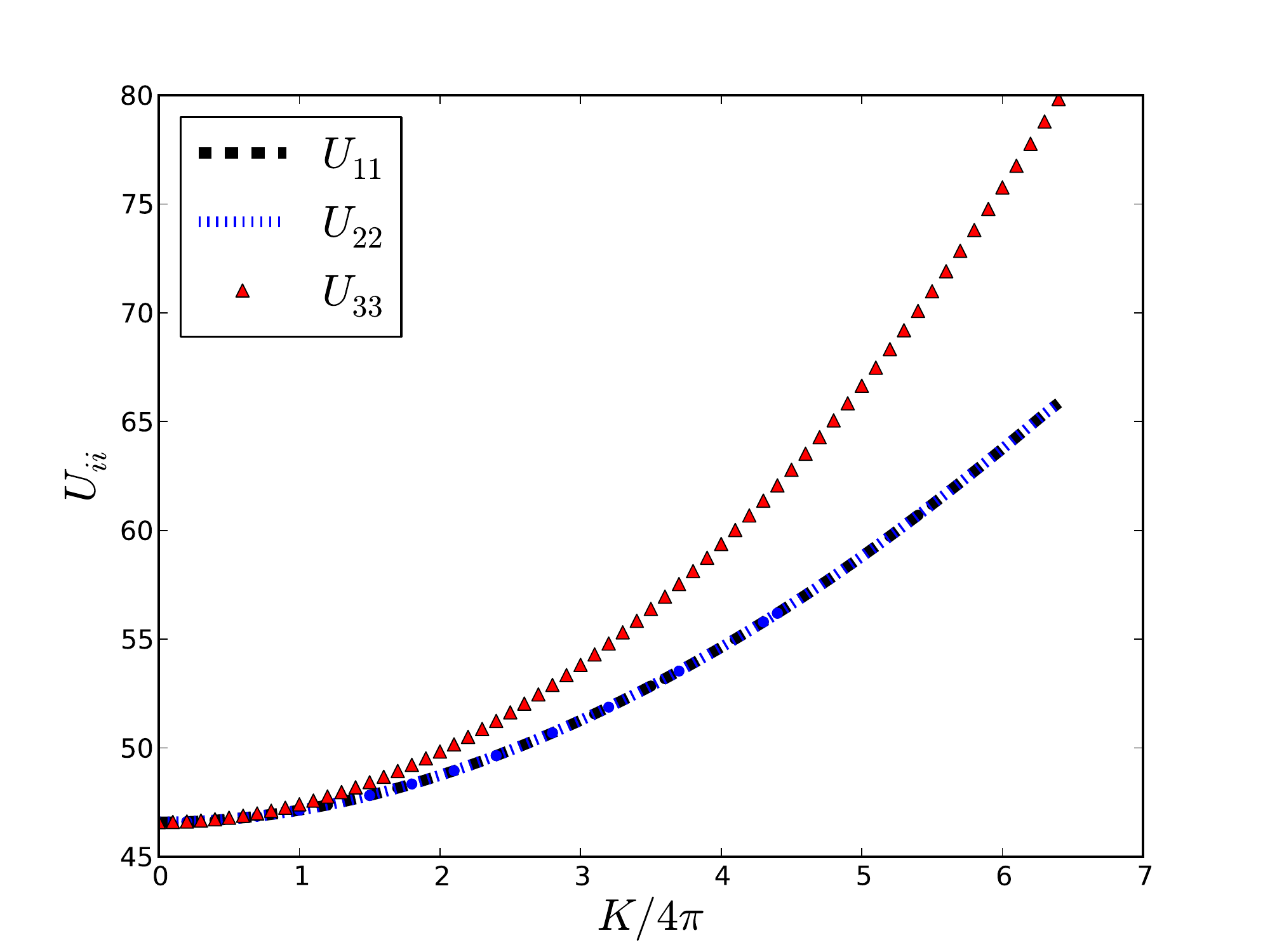}}
\caption[Total energy and inertia as function of $\omega$ and $K$ for isospinning $B=1$ Skyrmions.]{Isospinning $B=1$ Skyrmion $(\mu=1)$. A suitable start configuration of topological charge $1$ is numerically minimized using a damped field evolution (3D modified Newtonian flow) on a $(200)^3$ grid with a lattice spacing of $\Delta x=0.1$ and a time step size $\Delta t=0.01$. We choose the $z$ axis as our isorotation axis. Our energy results are compared with those obtained assuming an axially-symmetric deforming $B=1$ Skyrme configuration. Furthermore, we include the energy curve for a rigidly isorotating Skyrme configuration.}
\label{Fig_B1_O3}
\end{figure}

For the $O(3)$ symmetric charge-1 Skyrmion solution, we choose $\boldsymbol{\widehat{K}}=(0,0,1)$ as isospin axis. We observe that isospinning $B=1$ solutions are well approximated by an axially symmetric ansatz \cite{Battye:2005nx,Houghton:2005iu,Fortier:2008yj} when allowing for centrifugal deformations. We display in Fig.~\ref{Fig_B1_O3}(a) the total energy $E_{\text{tot}}$ as a function of isospin $K$ for $B=1$ Skyrmions of mass value $\mu=1$ and calculated without imposing any symmetry constraints on the isospinning Skyrme configuration. In our 3D simulations the soliton's energy is given by $E_\text{tot}(\omega)=M_1+K^2/2U_{33}$ and the energy curve terminates at $K_{\text{crit}}= 6.5\times 4\pi$ ($\omega_{\text{crit}} = \mu = 1$). Stable, internally spinnning solutions cease to exist beyond this critical value, but energy and moments of inertia remain finite at $K_{\text{crit}}= 6.5\times 4\pi$. We find that the energy curve $E_{\text{tot}}(K)$ matches within the limits of our numerical accuracy the energies of isospinning $B=1$ Skyrmions when assuming an axial symmetry. 

 As shown in Fig.~\ref{Fig_B1_O3}(a), the rigid body formula proves to be a good approximation for small isospins ($K\le 3\times 4\pi$), whereas for higher isospin values $E_{\text{tot}}(K)$ deviates from the quadratic behavior. At the critical angular frequency $\omega_{\text{crit}}=1$ ($K_{\text{crit}}= 6.5\times 4\pi$) the rigid body approximation gives an approximate  $7\%$ larger energy value for the isospinning soliton solution. The associated isospin inertia tensor $U_{ij}$ (\ref{U_phi_not}) is diagonal and its diagonal elements as a function of isospin $K$ are shown in Fig.~\ref{Fig_B1_O3}(b). For small isospin values the Skyrme configuration possesses, within our numerical accuracy, $O(3)$ symmetry ($U_{ij}=V_{ij}=W_{ij}=\Lambda\delta_{ij}$ where the moment of inertia $\Lambda$ is calculated to be $47.5$) and as $K$ increases the soliton solution deforms by breaking the spherical symmetry to an axial symmetry (the tensors of inertia (\ref{Sky_Inertia}) are all diagonal and satisfy $U_{11}=U_{22}=u$, $V_{11}=V_{22}=v$, $W_{11}=W_{22}=w$ and $U_{33}=V_{33}=W_{33}$). At the critical angular frequency $\omega_{\text{crit}}=1$ ($K_{\text{crit}}= 6.5\times 4\pi$) we find numerically $u= 66.7$, $v=67.9$,  $w=58.52$  and $U_{33}= 80.9$. 

\begin{figure}[!htb]
\subfigure[\,$B=1$, $\boldsymbol{\widehat{K}}=(0,0,1)$]{\includegraphics[totalheight=6.cm]{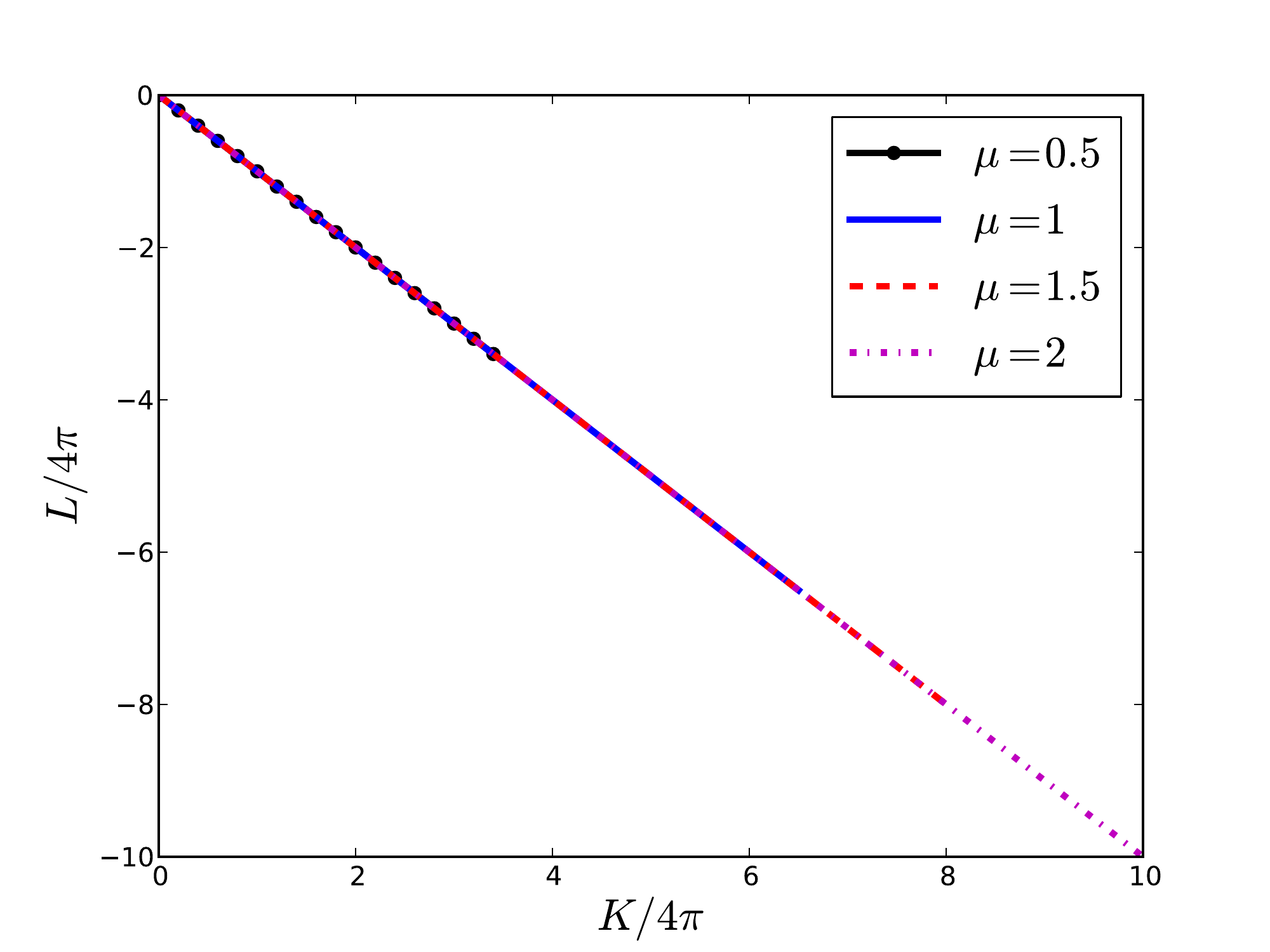}}
\subfigure[\,$B=2$, $\boldsymbol{\widehat{K}}=(0,0,1)$]{\includegraphics[totalheight=6.cm]{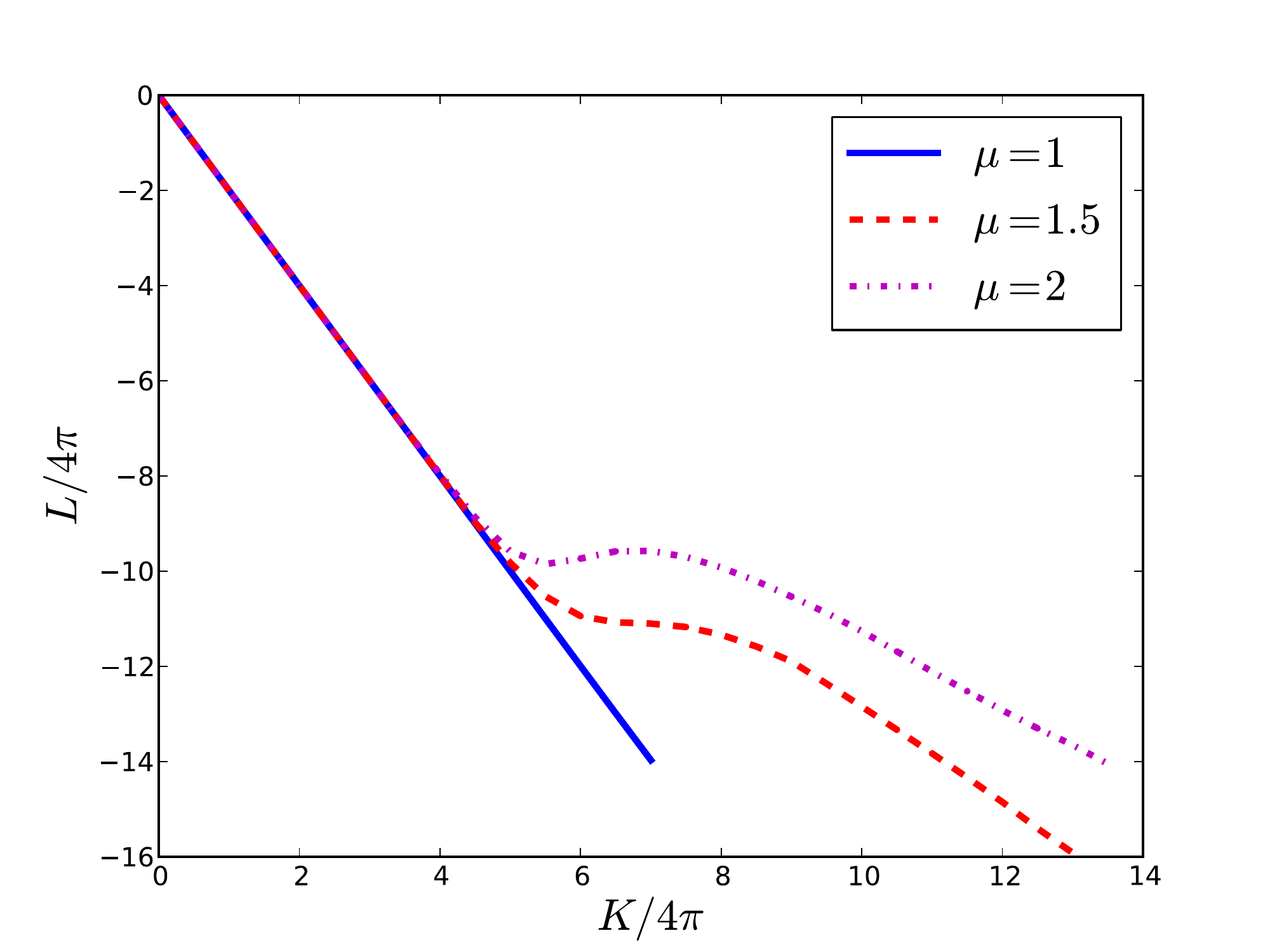}}\\
\subfigure[\,$B=3$, $\boldsymbol{\widehat{K}}=(0,0,1)$]{\includegraphics[totalheight=6.cm]{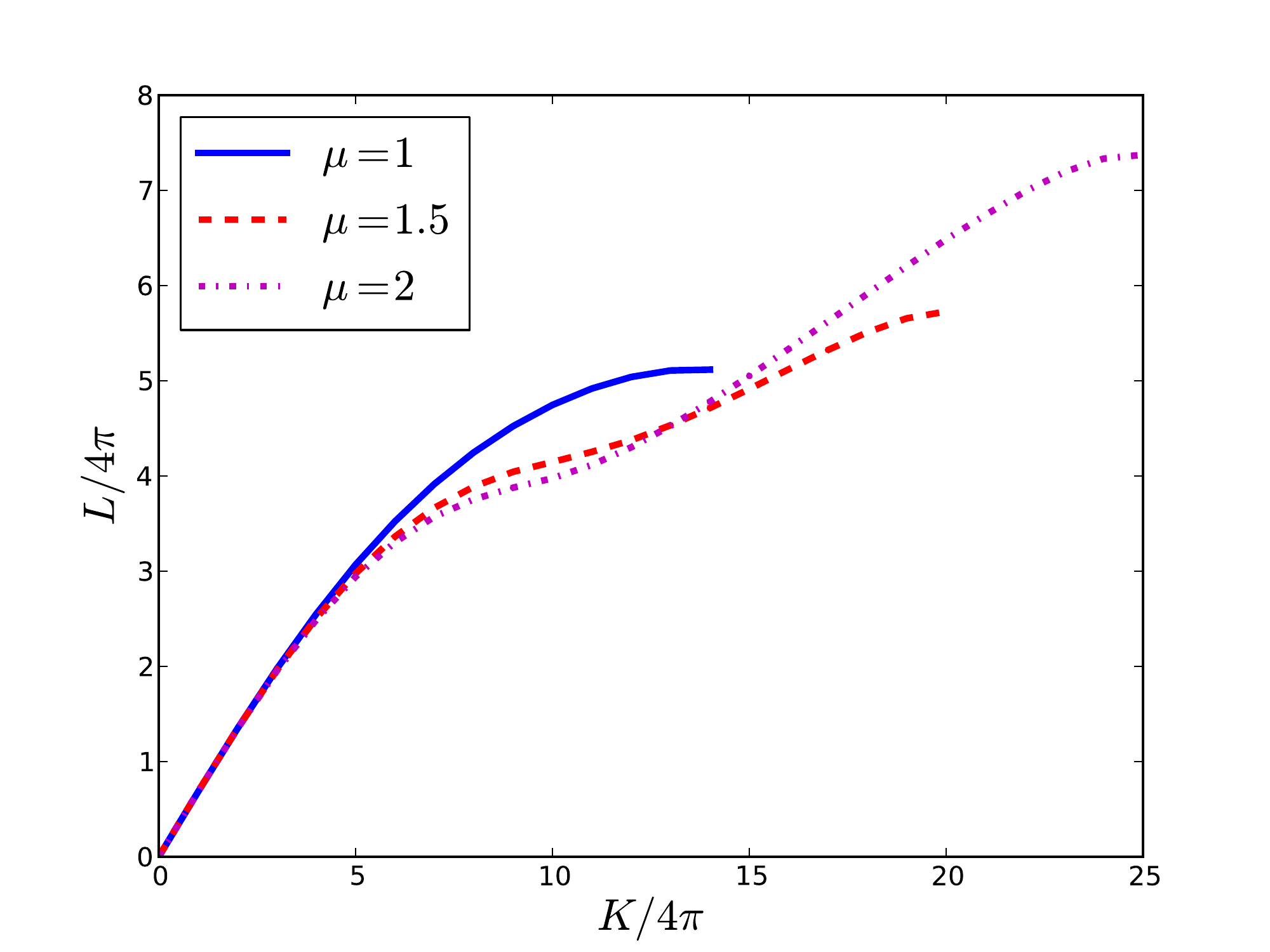}}
\subfigure[\,$B=8$ ($D_{6d}$ symmetry), $\boldsymbol{\widehat{K}}=(0,0,1)$]{\includegraphics[totalheight=6.cm]{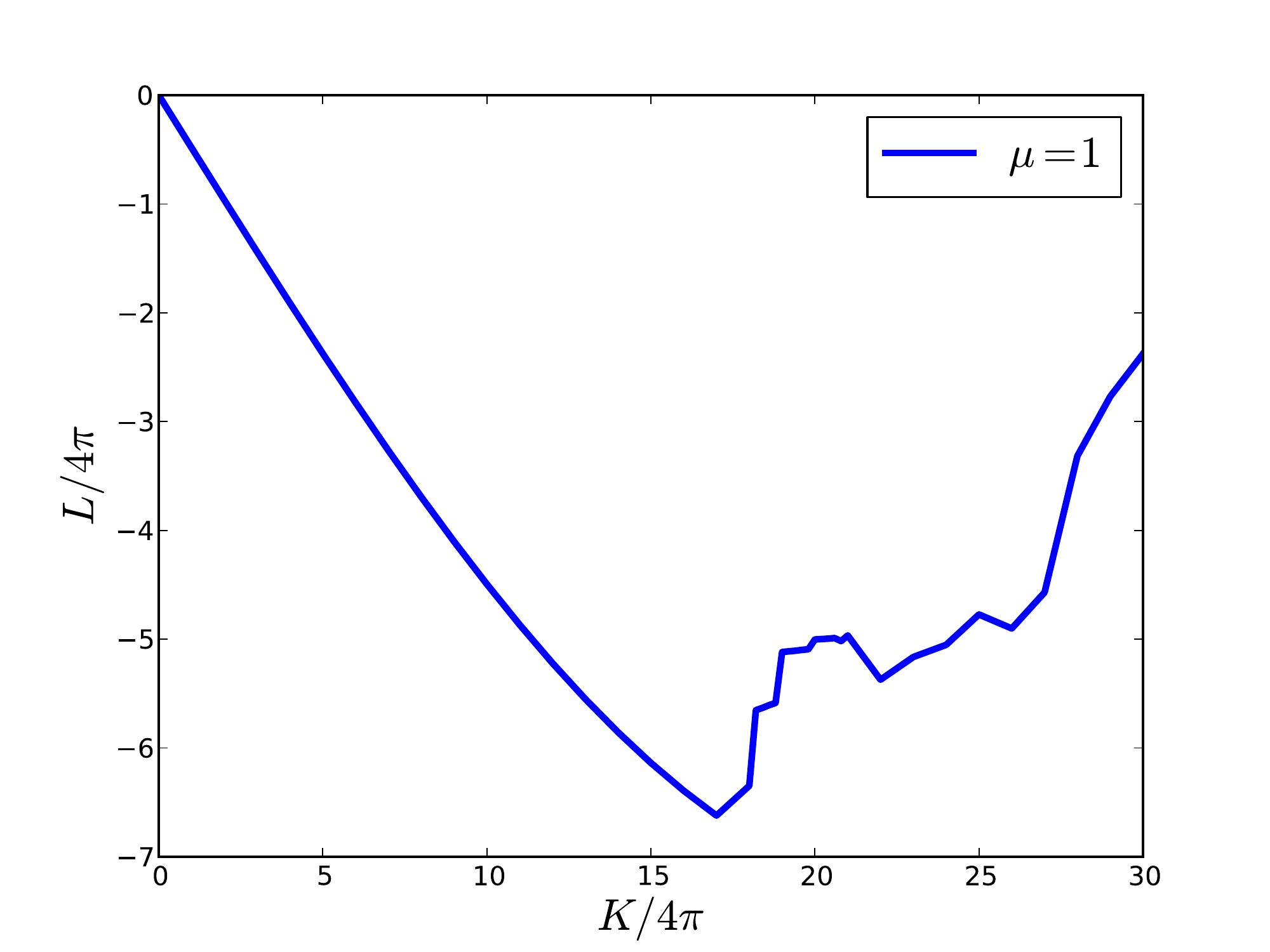}}
\caption{Spin $L$ as a function of isospin $K$ for Skyrmion solutions of topological charge $B=1-3,8$ and with pion mass $\mu$. The isospin axes are chosen as indicated. Here, we only display isospinning Skyrme configurations with nonzero $L$. }
\label{B_Spin}
\end{figure}

As pointed out in Section~\ref{Sec_Sky_Iso}, Skyrme configurations can acquire classical spin if the mixed inertia tensor $W_{ij}$ is nonzero. For $B=1$ we display in Fig.~\ref{B_Spin}(a) the acquired spin $L$ as a function of isospin $K$ for a range of pion masses $\mu$ when isospinning about $\boldsymbol{\widehat{K}}=(0,0,1)$. Since axial symmetry is preserved, $L$ grows linearly with $K$.    Numerically, the slope is found to be $-0.99$ and agrees well with the expected one $L/K=-W_{33}/U_{33}=-1$ for an axially symmetric charge-1 configuration with $W_{33}=U_{33}$. Stationary, isospinning Skyrme configurations can only be constructed up to the critical isospin $K_{\text{crit}}=U_{33}\omega_{\text{crit}}$ with $\omega_{\text{crit}}=\mu$. Consequently, the functions $L(K)$ terminate at the points $L_{\text{crit}}=-K_{\text{crit}}$. Spin $L$ and isospin $K$ have the same magnitude and are of opposite sign. This agrees with the  Finkelstein-Rubinstein constraints that are commonly imposed when quantizing $B=1$ Skyrmions \cite{Krusch:2002by} . 

\subsubsection{$B=2$}

For $B=2$, we investigate isospinning Skyrmion solutions with isospin axes $\boldsymbol{\widehat{K}}=(0,1,0)$ and $\boldsymbol{\widehat{K}}=(0,0,1)$. Isospinning around $\boldsymbol{\widehat{K}}=(0,1,0)$ leads to a novel configuration with $D_4$ symmetry  and isospinning around $\boldsymbol{\widehat{K}}=(0,0,1)$ can result into the break up into two charge-1 Skyrmions [see baryon density isosurfaces displayed in Fig.~\ref{Fig_B1B8_Sky}(a) and \ref{Fig_B1B8_Sky}(b)]. Note that for fixed isospin the $D_4$-symmetric Skyrme configuration has lower energy than the one formed by two charge-1 Skyrmions and isospinning around $\boldsymbol{\widehat{K}}=(0,0,1)$. Indeed, we observe that when perturbing a $B=2$ Skyrmion isospinning  around $\boldsymbol{\widehat{K}}=(0,0,1)$ the configuration can relax into the lower energy $B=2$ configuration isospinning around $\boldsymbol{\widehat{K}}=(0,1,0)$.  For the $D_4$-symmetric  configuration, we verify numerically that the spin $L=-W_{22}\omega$ vanishes for all classically allowed isospin values $K$.  Therefore, this configuration may become important for calculating excited states of the Deuteron with nonzero isospin \cite{Manton:2011mi}.

However, if we choose $\boldsymbol{\widehat{K}}=(0,0,1)$ as isospin axis, the $B=2$ Skyrmion gains spin $L=-W_{33} K$ as $K$ increases, see Fig.~\ref{B_Spin}(b).  For $\mu=1$ the  axial symmetry remains unbroken and hence $L$ depends linearly on $K$. We confirm that the numerically calculated slope $-1.97$ agrees well with the expected one $L/K=-W_{33}/U_{33}=-2$ as $W_{33}=2U_{33}$ for an axially symmetric   charge-2 configuration. For larger mass values $\mu$  isospinning around the
$\boldsymbol{\widehat{K}}=(0,0,1)$ leads to the break up into two $B=1$ Skyrmions orientated in the attractive channel. The axial symmetry is broken at $K_{\text{SB}}$, and the $B=2$ Skyrmion solution starts to split into two $B=1$ Skyrmions.  As the isospin $K$ increases further $L$ is approximately $-K$. In this regime the isospinning configuration is well described by two separated, axially symmetric deformed $B=1$ Skyrmions.  This is consistent with head-on scattering of two spinning $B=1$ Skyrmions in the attractive channel \cite{DFosterandSKruschinpreparation} where the configuration of closest approach is not the torus but a configuration of two separated Skyrmions. The attractive channel has also been discussed in Ref.~\cite{Leese:1994hb} when quantising the Deuteron. This degree of freedom was essential for comparing the spatial probability distribution of the deuteron with experimental values \cite{Manton:2011mi,Braaten:1988cc,Forest:1996kp}.

In summary, for $B=2$ we observe that Skyrme configurations with nonzero $W$ and hence nonzero spin $L$ show centrifugal effects and separate out whereas states with $W=0$ tend to stay more compact. Isospinning the charge-2 Skyrmion about its $(0,0,1)$ axis results in the breakup into two well separated charge-1 Skyrmions, whereas isospinning about $(0,1,0)$ yields compact $D_4$-symmetric $B=2$ configurations of lower energy.

\subsubsection{$B=3$}

For $B=3$, we isorotate the minimal-energy tetrahedron about its $\boldsymbol{\widehat{K}}=(0,0,1)$ axis. 
For mass value $\mu=1.5$ and $K$ sufficiently large, we observe that the isospinning $B=3$ Skyrmion forms a distorted ``pretzel\,'' configuration -- a state that has previously been found to be meta-stable \cite{Walet:1996he,Battye:1996nt} for vanishing isospin $K$ [see baryon density isosurfaces in Fig.~\ref{Fig_B1B8_Sky}(c)]. For higher pion masses $\mu$,  we find that the tetrahedral charge-3 Skyrmion even can break into lower-charge Skyrmions as the angular frequency $\omega$ increases. For pion mass $\mu=2$ the isospinning charge-3 Skyrmion solution seems to pass through the distorted ``pretzel\,'' configuration as $\omega$ increases. Then it breaks into a toroidal $B=2$ Skyrmion solution and a $B=1$ Skyrmion before reaching its upper frequency limit $\omega_\text{crit}=\mu$. 

For $B=3$ we display in Fig.~\ref{B_Spin}(c) the $L(K)$ graphs for pion masses $\mu=1,1.5,2$. We observe that as long as the tetrahedral symmetry remains unbroken the spin $L$ inreases linearly with $K$. Breaking of the tetrahedral symmetry results in a lower $W$ and hence a lower increase in $L$ for higher $K$ values.

\subsubsection{$B=4$}

For $B=4$ there are two different isospin axes: $\boldsymbol{\widehat{K}}=(0,0,1)$ and $\boldsymbol{\widehat{K}}=(0,1,0)$.  For 
$\boldsymbol{\widehat{K}}=(0,0,1)$ we find that the octahedral symmetry remains unbroken [see baryon density isosurfaces shown in Fig.~\ref{Fig_B1B8_Sky}(d)]. When choosing $\boldsymbol{\widehat{K}}=(1,0,0)$ or $\boldsymbol{\widehat{K}}=(0,1,0)$ as our isorotation axis, we observe that with increasing angular frequency $\omega$ the octahedrally-symmetric charge-4 Skyrmion solution becomes unstable to break up into a pair of toroidal $B=2$ Skyrmions [see Fig.~\ref{Fig_B1B8_Sky}(e)]. Similar to the $B=2$ case, we find that the Skyrmion configuration of lowest energy for given isospin $K$ is the solution for which the constituents stay closer together, that is the $B=4$ cube isospinning about  $\boldsymbol{\widehat{K}}=(0,0,1)$. The charge-4 solutions which split into two $D_4$-symmetric charge-2 tori are of higher energy for all classically allowed isospin values $K$. For both axis choices the mixed inertia tensor $W$ and hence $L$ vanish for all classically allowed isospin values $K$.

\subsection{Higher Charge Skyrmions ($B=8$)}

For $B=8$ it is interesting to observe that all configurations are breaking up into constituents, either into two $B=4$ parts or into four $B=2$ parts, see Fig.~\ref{Fig_B8_Sky}. Note that physically $B=8$ may describe beryllium $^8{\text{Be}}$ which is unstable to splitting up into two $\alpha$ particles, see e.g. Ref.~\cite{Krane:1987}. For $D_{6d}$-symmetric Skyrmion solutions we find that when isospinning about $\boldsymbol{\widehat{K}}=(0,0,1)$ there exist a critical isospin value at which the soliton solution splits up into two $B=4$ cubes. This breakup process is reflected in the $L(K)$ graph shown in Fig.~\ref{B_Spin}(d). For $K\le 17\times 4\pi$ ($L\ge -6.6 \times 4 \pi$) the spin $|L|$ grows linearly with $K$ and  the isospinning Skyrme configuration preserves its dihedral symmetry. For higher isospin values $K$ the dihedral symmetry is broken and the isospinning solution starts to break apart into two $B=4$ cubes of zero total spin $L$. Fig.~\ref{B_Spin}(d) shows how $L(K)$ decreases as $K$ increases beyond  $K=17\times 4\pi$.  When choosing $\boldsymbol{\widehat{K}}=(0,1,0)$ as isorotation axis, the  isospinning $D_{6d}$-symmetric Skyrmion solutions breaks up into four $B=2$ tori. In this case, the total spin $L$ is found to be zero (within the limits of our numerical accuracy) for all classically allowed isospin values. For $B=8$ Skyrme solitons with approximate $D_{4h}$ symmetry we find that all the isosponning solutions investigated in this article (see baryon density isosurfaces in Fig.~\ref{Fig_B8_Sky})  possess zero total spin for all values of $K$.

\section{Conclusions}\label{Sec_Con}
In this talk we summarized our results on isospinning soliton solutions with topological charges $B=1-4,8$ in the Skyrme model with the conventional mass term included and without imposing any assumptions about the soliton's spatial symmetries. Our numerical calculations show that the qualitative shape of isospinning Skyrmion solutions can differ drastically from the ones of the static ($\omega=0$) solitons. The deformations become increasingly pronounced as the mass value $\mu$ increases. Briefly summarized, we distinguish the following types of behavior:

\begin{enumerate}[(i)]
\item \emph{Breakup into lower charge Skyrmions:} Isospinning Skyrmion solutions can split into lower charge Skyrmions at some critical breakup frequency value. Examples are the breakup (for $\mu$ sufficiently large) of the $B=2$ solution into two $B=1$ Skyrmions when isospinning about $\boldsymbol{\widehat{K}}=(0,0,1)$; the breakup of the $D_{4h}$- and $D_{6d}$-symmetric $B=8$ Skyrme configurations into four $B=2$ tori when isospinning about $\boldsymbol{\widehat{K}}=(0,1,0)$ and the breakup of isospinning $D_{4h}$ and $D_{6d}$ Skyrmions into charge-4 subunits. 
\item \emph{Formation of new solution types:}  Isospinning Skyrmion solutions can deform into configurations that do not exist at vanishing $\omega$ or are only metastable at $\omega=0$. An example is the tetrahedral $B=3$ Skyrmion (with $\mu=1.5$) which evolves with increasing $\omega$ into a ``pretzel\,''-like configuration -- a state that is only metastable at $\omega=0$ \cite{Walet:1996he,Battye:1996nt}.  
 \item \emph{Lifting of energy degeneracies:} Adding isospin can remove energy degeneracies. For example isospinning $D_{6d}$- and $D_{4h}$-symmetric Skyrme solitons about their $\boldsymbol{\widehat{K}}=(0,0,1)$ axes results for the configuration with approximate $D_{4h}$ symmetry in a lower energy value than found for isospinning $D_{6d}$ solitons, thereby removing the degeneracy.
\item \emph{Spin generated from Isospin:} 
If $W_{ij}$ is nonzero, then Skyrme configurations will obtain classical spin when isospin is added. For example for $B=1$ this gives states with spin and isospin opposite, as required by the Finkelstein-Rubinstein constraints \cite{Krusch:2002by}. For $B=2$ we observe that Skyrme configurations with nonzero mixed inertia tensor $W_{ij}$ show centrifugal effects and separate out whereas states with $W_{ij}=0$ tend to stay more compact. Isospinning around 
$\boldsymbol{\widehat{K}}=(0,0,1)$ leads to the breakup into two $B=1$ Skyrmions orientated in the
attractive channel and with $L_{\text{crit}}$ given by approximately $-K_{\text{crit}}$. Isospinning around $\boldsymbol{\widehat{K}}=(0,1,0)$ leads to a novel configuration with $D_4$ symmetry of vanishing total spin for all classically allowed isospin values $K$. 
\end{enumerate}

The types of deformations presented in this talk have been largely ignored in previous work \cite{Manko:2007pr,Battye:2009ad} on modelling nuclei by quantized Skyrmion solutions and are exactly the ones we would like to take into account when quantizing the Skyrme model. Spin and isospin quantum numbers of ground states and excited states have so far almost exclusively been calculated within the rigid body approach \cite{Irwin:1998bs,Krusch:2002by,Krusch:2005iq,Lau:2014sva}, that is by neglecting any deformations and symmetry changes due to centrifugal effects. Our numerical full field simulations clearly demonstrate the limitations of this simplification. The symmetries of isospinning soliton solutions can change drastically and the solitons are found to be of substantially lower energies than predicted by the rigid body approach. This work offers interesting new insights into the classical behavior of Skyrmions and gives an indication of which effects have to be taken into account when quantising Skyrmions.

\section*{Acknowledgements}
We would like to thank the organizers of Quarks-2014 for a very interesting conference in Suzdal. This work was undertaken on the COSMOS Shared Memory system at DAMTP, University of Cambridge operated on behalf of the STFC DiRAC HPC Facility. This equipment is funded by BIS National E-infrastructure capital grant ST/J005673/1 and STFC grants ST/J001341/1, ST/H008586/1, ST/K00333X/1. The work presented in this talk is based on the recent article \cite{Battye:2014qva} and was done in collaboration with Richard Battye and Steffen Krusch. Many thanks to David Foster, Nick Manton, Yasha Shnir, Paul Sutcliffe and Niels Walet for useful discussions.
This work was financially supported by the UK Engineering and Physical Science Research
Council (grant number EP/I034491/1).

\small{

\providecommand{\href}[2]{#2}\begingroup\raggedright\endgroup}

\end{document}